\documentclass{article}
\usepackage{isit2004}  % Uncomment this line for LaTeX2e
\begin{document}
%\itwAfour             % Uncomment for A4 paper
\topmargin = 0mm

%\itwtitle{Quantum Key Distribution with Vacua
%\\[0.5ex]
%  or Dim Pulses as Decoy States}
\itwtitle{Quantum Key Distribution with Vacua or Dim Pulses as Decoy States}

%\\[0.5ex]
%  or Dim Pulses as Decoy States}

%--- Style I ----------------------------------------------------------------
%  For an author, the macro
%  \itwauthorA may be used instead. (The three styles cannot be mixed.)

%\itwauthorA{Hoi-Kwong Lo\footnotemark[1]}
%          {Dept. of Electrical \& Computer Engineering \& \\
%Dept. of Physics, \\
%University of Toronto,\\
%10 King's College Road, \\
%Toronto, Ontario, CANADA, M5S 3G4 \\
%e-mail: {\tt hklo@comm.utoronto.ca }}

\itwauthorA{Hoi-Kwong Lo\footnotemark[1]}
          {Dept. of Electrical \& Computer Engineering \& \\
Dept. of Physics, University of Toronto, 10 King's College Road, \\
Toronto, Ontario, CANADA, M5S 3G4, e-mail: {\tt hklo@comm.utoronto.ca }}

\itwmaketitle

\footnotetext[1]{This work was supported in part by
Canadian NSERC, Canada Research Chairs Program, Connaught Fund,
Canadian Foundation for Innovations, Ontario Innovation Trust,
Premier's Research Excellence Award, Canadian Institute for
Photonics Innovations, MagiQ Technologies, Inc., New York, and
University of Toronto start-up grant.}

\begin{itwabstract}
    Recently, Hwang has proposed a decoy state method in quantum key
distribution (QKD). In Hwang's proposal,
the average photon number of the decoy state is about two.
Here, we propose a new decoy state scheme using vacua or
very weak coherent states as decoy states and discuss its
advantages.

\end{itwabstract}

\begin{itwpaper}

\itwsection{Introduction}

The security of conventional cryptography is often based on
unproven computational assumptions such as the hardness of factoring.
In contrast, quantum cryptography offers unconditional security based
on fundamental laws of physics, particularly, the quantum no-cloning
theorem.
The best-known application of quantum cryptography
is quantum key distribution
(QKD). The goal of QKD is to
allow two users to communicate in absolute security in the presence of
an eavesdropper. The best-known QKD protocol (BB84) was published
by Bennett and Brassard in 1984 \cite{BB84}.

The procedure of standard BB84 QKD scheme is as follows.
First, Alice sends Bob
a time-ordered sequence of single photons, each of which in
one of the four possible polarizations---horizontal, vertical,
45-degrees and 135-degrees.
Second, for each photon, Bob makes randomly one of two measurements:
diagonal or rectilinear.
Third,
Bob writes down the bases and measurement outcomes
and publicly acknowledges his receipt of
Alice's photons. Fourth, Alice and Bob then broadcast their bases.
For the cases when they use different bases,
they throw away their polarization data.
For the cases when they use the same bases,
they keep their polarization data.
Note that in the absence of noise and Eve, Alice
and Bob's polarization data should be the same.
Next, Alice and Bob perform some test for tampering.
For instance, Alice and Bob can pick a random subset of
their photons and broadcast their polarization data. From
there, they compute the quantum bit error rate (QBER) of their
transmission. If this QBER is larger than some prescribed value,
they abort the protocol. On the other hand, if this QBER is within
some prescribed value, then the QKD protocol is successful
and they proceed with key generation.
Key generation may be done by applying two different operations:
a) error correction and b) privacy amplification.

The security of QKD protocols with ideal devices
and, more recently, with imperfect devices has been proven.
Since a single-photon source is an experimental challenge,
a common experimental source is a weak (coherent) laser pulse,
which, if phase randomized, can be regarded as a Poisson
distribution in the number of photons.
This means that there is a non-zero probability of having
multi-photons. In principle, Eve can perform a quantum
non-demolition measurement in photon number.
If she sees a single photon, she suppresses the signal.
If she sees a multi-photon, she steals one photon and resends
the rest to Bob. She may try to mask her presence by the
usual loss in a quantum channel. Note, however, that with
such an attack, multi-photon signals exhibit different
outcomes from single-photon signals. For instance,
multi-photons give a much higher yield than single photons.

Recently, Hwang \cite{hwang} has provided an ingenious method to defend
against the above attack by Eve. His idea is that,
in addition to regular signal states, Alice also uses
some decoy states. For each signal, Alice randomly chooses to
send either a signal state or a decoy state. After Bob's
measurements of all signals, Alice tells Bob which signals
are decoy states. They can then compare their outcomes for the
decoy states and use their data to detect eavesdropper's
attack.

In Hwang's original proposal, the average photon number of the
decoy state is about two and is,
thus, rather high by QKD standard. The goal there is to use the observed
fractional loss of the decoy state to constrain the multi-photon
yield in QKD, thus detecting eavesdropping attacks.

\itwsection{Our idea}

We propose several new ideas here.
First of all, we propose a refined data analysis
where the specific properties of the
decoy state such as the bit error rate are analyzed separately from
the original signal.
Second, we show that Hwang's scheme can be combined with existing
security proofs of QKD \cite{GLLP,twoway}, thus
allowing rigorous proofs of security.
Third, we show that it may be advantageous to use either vacua or very
weak coherent states or both as decoy states. On one hand,
by using a vacuum as a decoy state, Alice and Bob can verify the
so-called dark
count rates of their detectors.
On the other hand, by using a very weak coherent pulse as
a decoy state, Alice and Bob can easily lower
bound the yield of single-photon pulses.
These two types of decoy
states can be easily combined with each other and with
Hwang's original proposal
The usage of decoy states may have some advantages. For instance,
they may provide useful testing procedure and may increase the secure
key generation rate, increase security and the distance of QKD.
One specific example is that they can achieve secure QKD even at high loss.

\begin{itwreferences}

\bibitem{BB84} C. H. Bennett and G. Brassard, Quantum cryptography:
            Public key distribution and coin tossing, {\it Proceedings of
            IEEE International Conference on Computers, Systems, and Signal
            Processing},  IEEE, 1984, pp. 175-179.

\bibitem{hwang} W.-Y. Hwang, Phys. Rev. Lett. {\bf 91}, 057901 (2003).

\bibitem{GLLP} D. Gottesman, H.-K. Lo, N. L\"utkenhaus, and J. Preskill,
Security of quantum key distribution with imperfect Devices,
on-line available at
http://xxx.lanl.gov/abs/quant-ph/0212066

\bibitem{twoway} D. Gottesman and H.-K. Lo, IEEE Transactions on
Information Theory {\bf 49}, No. 2, p. 457 (2003).

\end{itwreferences}
\end{itwpaper}
\end{document}